\documentclass[prb,twocolumn,showpacs,preprintnumbers,amsmath,amssymb]{revtex4}
\usepackage{graphicx}% Include figure files
\usepackage{dcolumn}% Align table columns on decimal point
\usepackage{bm}% bold math

\begin{document}

\newcommand*{\cm}{cm$^{-1}$\,}
\newcommand*{\KFeSe}{K$_x$Fe$_{2-y}$Se$_2$\,}
\newcommand*{\Tc}{T$_c$\,}

%\reprint{APS/123-QED}

\title{Peculiar structures in the infrared spectra of new iron selenide K$_{0.83}$Fe$_{1.53}$Se$_2$:
implication for the electronic structure}% Force line breaks with \\
\author{Z. G. Chen}
\author{R. H. Yuan}
\author{T. Dong}
\author{G. Xu}
\author{Y. G. Shi}
\author{P. Zheng}
\author{J. L. Luo}
\author{J. G. Guo}
\author{X. L. Chen}
\author{N. L. Wang}
\affiliation{Beijing National Laboratory for Condensed Matter
Physics, Institute of Physics, Chinese Academy of Sciences,
Beijing 100190, China}

%\date{March 26, 2008}% It is always \today, today,

\begin{abstract}
We report an infrared spectroscopy study on
K$_{0.83}$Fe$_{1.53}$Se$_2$, a semiconducting parent compound of
the new iron-selenide system. The major spectral features are
found to be distinctly different from all other Fe-based
superconducting systems. Our measurement revealed two peculiar
spectral structures: a double peak structure between 4000-6000
cm$^{-1}$ and abundant phonon modes much more than those expected
for a 122 structure. We elaborate that those features could be
naturally explained from the blocked antiferromagnetism due to the
presence of Fe vacancy ordering as determined by recent neutron
diffraction experiments. The double peaks reflect the coexistence
of ferromagnetic and antiferromagnetic couplings between the
neighboring Fe sites.
\end{abstract}

\pacs{74.70.Xa, 74.25.Gz, 74.25.nd}


\maketitle

After three years intensive studies on the Fe-based
superconductors, much progress has been made in understanding the
properties and principles of materials. Similar to the high-T$_c$
cuprates, the superconductivity in Fe-based compounds is found to
be in close proximity to an antiferromagnetic (AFM)
order\cite{Dong,Cruz}. Superconductivity emerges when the magnetic
order was suppressed by electron or hole doping or application of
pressure\cite{Dong,Cruz,Chen1,Rotter2,Torikachvili}. Although the
phase diagram of Fe-pnictides appears very similar to that of
high-$T_c$ curpates, distinct differences exist between them. The
undoped compounds in cuprates are Mott insulators, by contrast,
the parent compounds in Fe-pnictides are spin-density-wave (SDW)
metals. Electronic structure studies show that the Fermi surfaces
and band structures of Fe-based compounds are qualitatively
similar. In particular, all the compounds show small compensating
hole and electron Fermi surfaces locating respectively near the
Brillouin zone center and corner\cite{Singh1}. It is widely
believed that the inter-pocket scattering between the electron and
hole pockets is crucial to the superconducting
pairing\cite{Kuroki,Mazin,Christianson,Richard,Wang1,Eremin}.

However, the recent discovery of superconductivity over 30 K in
K$_x$Fe$_2$Se$_2$ \cite{Guo} poses a strong challenge to the above
picture. Soon after this discovery, a resistivity study on a
related system (Tl,K)Fe$_{2-x}$Se$_2$ indicated that the
superconducting phase evolves from an insulating phase rather than
an SDW metal \cite{Fang}. Subsequent angle resolved photoemission
spectroscopy (ARPES) studies revealed that the Fermi surface (FS)
topologies of those compounds are very different from previously
known materials. Only the electron pockets are present in the
superconducting compounds, while the hole bands sink below the
Fermi level, indicating that the inter-pocket scattering between
the hole and electron pockets is not an essential ingredient for
superconductivity \cite{Zhang1,Qian}. More surprisingly, recent
muon-spin relaxation ($\mu$SR) \cite{Shermadini,Pomjakushin},
neutron diffraction \cite{Bao1,FYe,Bao2}, resistivity and
magnetization\cite{Liu} measurements on A$_x$Fe$_{2-y}$Se$_2$
(A=K, Rb, Cs, Tl) revealed a coexistence of superconductivity with
a strong AFM order. In particular, the neutron experiments
revealed a completely new type of magnetic order, a blocked
checkerboard antiferromagnetism for the system. The ordered
magnetic moment $\sim$ 3.31 $\mu_B$/Fe is unprecedentedly large,
and the the magnetic transition occurs at a record high
temperature of $T_N$=559 K below the formation of an Fe vacancy
ordering with a $\sqrt{5}\times\sqrt{5}\times1$ superlattice
pattern at T$_S$$\sim$ 578 K.

Infrared spectroscopy is a powerful technique to investigate
charge dynamics and band structure of a material as it probes both
free carriers and interband excitations. In this paper, we report
an infrared spectroscopy study on the insulating parent compound
of K$_{0.83}$Fe$_{1.53}$Se$_2$. Our measurement revealed a small
energy gap in the low-lying excitation. Furthermore, a double peak
structure between 4000-6000 \cm and abundant phonon lines much
more than those expected for a standard 122 structure were
observed. Those peculiar features were not seen in any other
Fe-pnictides/chalcogenides, and could be taken as the
characteristic optical spectral structures for this system. We
elaborate that those features could be naturally explained from
the blocked antiferromagnetism due to the presence of Fe vacancy
ordering as determined by recent neutron diffraction experiments.
Those measurement results are important in understanding the new
Fe-selenide system.

The single crystal samples were grown from self-melting method
with a nominal starting composition of K:Fe:Se=1:2:2. FeSe was
firstly synthesized by reacting Fe powder and Se powder at
700$^0$C for 15 hours twice. K pieces and FeSe powder were then
put into an alumina crucible and sealed in a Ta tube with Argon
gas at the pressure of 1 atom. The Ta tube was further sealed into
a quartz tube in vacuum. The crystal growth was taken place in a
box furnace. The crucible was slowly heated up to 1027$^0$C and
held for 5 hours, then cooled down to 750$^0$C at a rate of
5$^0$C/hour. Plate-like single crystals with shiny surface were
obtained after breaking the crucible.

\begin{figure}
\includegraphics[width=8 cm]{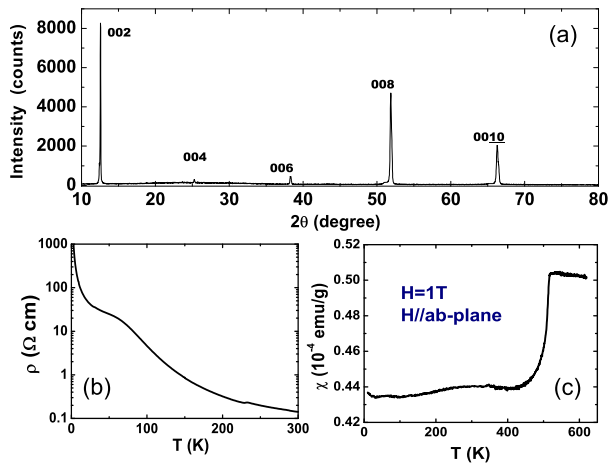}
\caption{(Color online) (a) The x-ray-diffraction pattern of the
crystal. Only \textit{(00l)} diffraction peaks are observed. (b)
The ab-plane resistivity curves versus temperature. (c) The
magnetic susceptibility curve versus temperature. The sharp drop
near 519 K indicates the development of the antiferromagnetic
order.}
\end{figure}

The crystals were characterized by X-ray diffraction (XRD),
scanning electron microscopy equipped with the energy dispersive
X-ray (EDX) spectroscopy, dc resistivity, and magnetic
susceptibility measurements. Figure 1 (a) shows the XRD pattern
for the single crystal with Cu K$\alpha$ radiation. As expected,
only \textit{(00l)} diffraction peaks are observed, indicating
that the crystallographic c axis is perpendicular to the cleaved
surface. The c-axis lattice constant is determined to be 14.10
$\AA$. The average composition determined by the EDX analysis on
several different positions of the crystal is found to be close to
K$_{0.83}$Fe$_{1.53}$Se$_2$. We use this composition throughout
the paper. However, we remark that this composition formula should
be considered only as an approximate one since the accuracy of EDX
analysis is not very high. Figure 1 (b) shows the in-plane dc
resistivity data measured by four-leads method in a Quantum Design
physical properties measurement system (PPMS). It shows an
insulating behavior. The resistivity increases by several orders
with decreasing temperature from 300 K. Figure 1 (c) shows the
magnetic susceptibility data below 620 K measured also in PPMS
under the filed of 1 Tesla. The sample shows a very sharp magnetic
transition at 518 K. Clearly, the crystal is an AFM insulator.

\begin{figure}
\includegraphics[width=6.5cm]{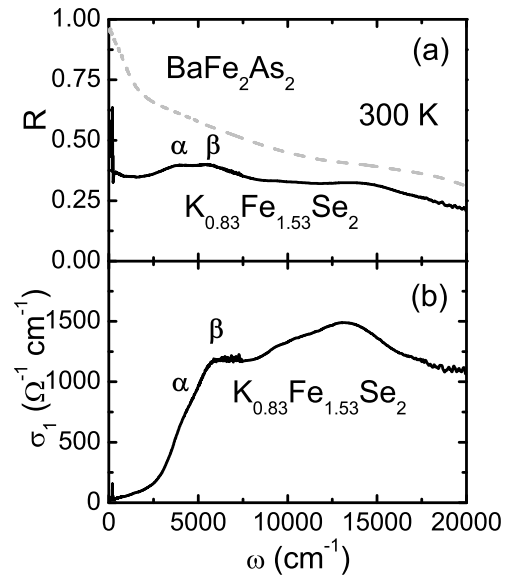}
\caption{The optical reflectance R($\omega$) (a) and conductivity
$\sigma_1(\omega)$ (b) at room temperature up to 20000 \cm.
$\alpha$ and $\beta$ indicate two-peaks interband transition
structure. The reflectance curve of BaFe$_2$As$_2$ at 300 K was
added for a comparison \cite{Hu122}.}
\end{figure}

Optical measurement was done on a Bruker Vertex 80v spectrometer
in the frequency range from 40 to 25000 cm$^{-1}$. The crystal
surface was found to be relatively stable, being comparable to
other Fe-pnictides/chalcogenides. The sample was mounted on an
optically black cone locating at the cold finger of the cryostat.
Freshly cleaved surface was obtained just before pumping the
cryostat. An \textit{in-situ} gold and aluminum overcoating
technique was used to get the reflectance \emph{R}($\omega$). The
optical data were found to be highly reproducible. The real part
of conductivity $\sigma_1(\omega)$ is obtained by the
Kramers-Kronig transformation of \emph{R}($\omega$). Figure 2
shows the R($\omega$) and $\sigma_1(\omega)$ spectra for the
sample over broad frequencies up to 20000 \cm. The overall
reflectance is rather low, roughly close to the value of 0.4. This
behavior is different from all other Fe-pnictide compound as well
as FeTe. As a comparison, we also plot the in-plane reflectance of
a BaFe$_2$As$_2$ crystal \cite{Hu122} in Fig. 2 (a) which is
considerably higher. The $\sigma_1(\omega)$ shows vanishing
conductivity at low frequency, indicating insulating-like
response. Detailed spectral features at low frequency will be
presented below. Strong interband transition peaks were seen near
5000 \cm and 13000 \cm. Actually the reflectance spectrum displays
two peaks (labelled as $\alpha$ and $\beta$) for the interband
transitions at lower energy scale, roughly at $\sim$ 4500 \cm
(0.56 eV) and 6000 \cm (0.75 eV). Relatively weak features are
seen in the $\sigma_1(\omega)$ spectrum.

Figure 3 shows the temperature dependence of the optical spectra
at lower energy scales. The left panels (3 (a) and (c)) show the
R($\omega$) and $\sigma_1(\omega)$ spectra up to 8000 \cm, the
right panels (3 (b) and (d)) show the spectra in the expanded low
frequency region within 400 \cm. Because the spectra display a
weak temperature dependence, we show the spectra only at three
different temperatures. The almost constant value of R($\omega$)
down to very low frequency yields evidence for a non-metallic
infrared response. The further decrease of the low frequency
R($\omega$) at lower temperature provides additional support for
the semiconducting or insulating behavior. Correspondingly, there
is no free carrier Drude response at low frequency in the
conductivity spectra (Fig. 3 (c) and (d)). The slight enhancement
of the low-$\omega$ optical conductivity at higher temperature can
be attributed to the thermally-activated hopping of electrons.

\begin{figure}
\includegraphics[clip,width=1.65in]{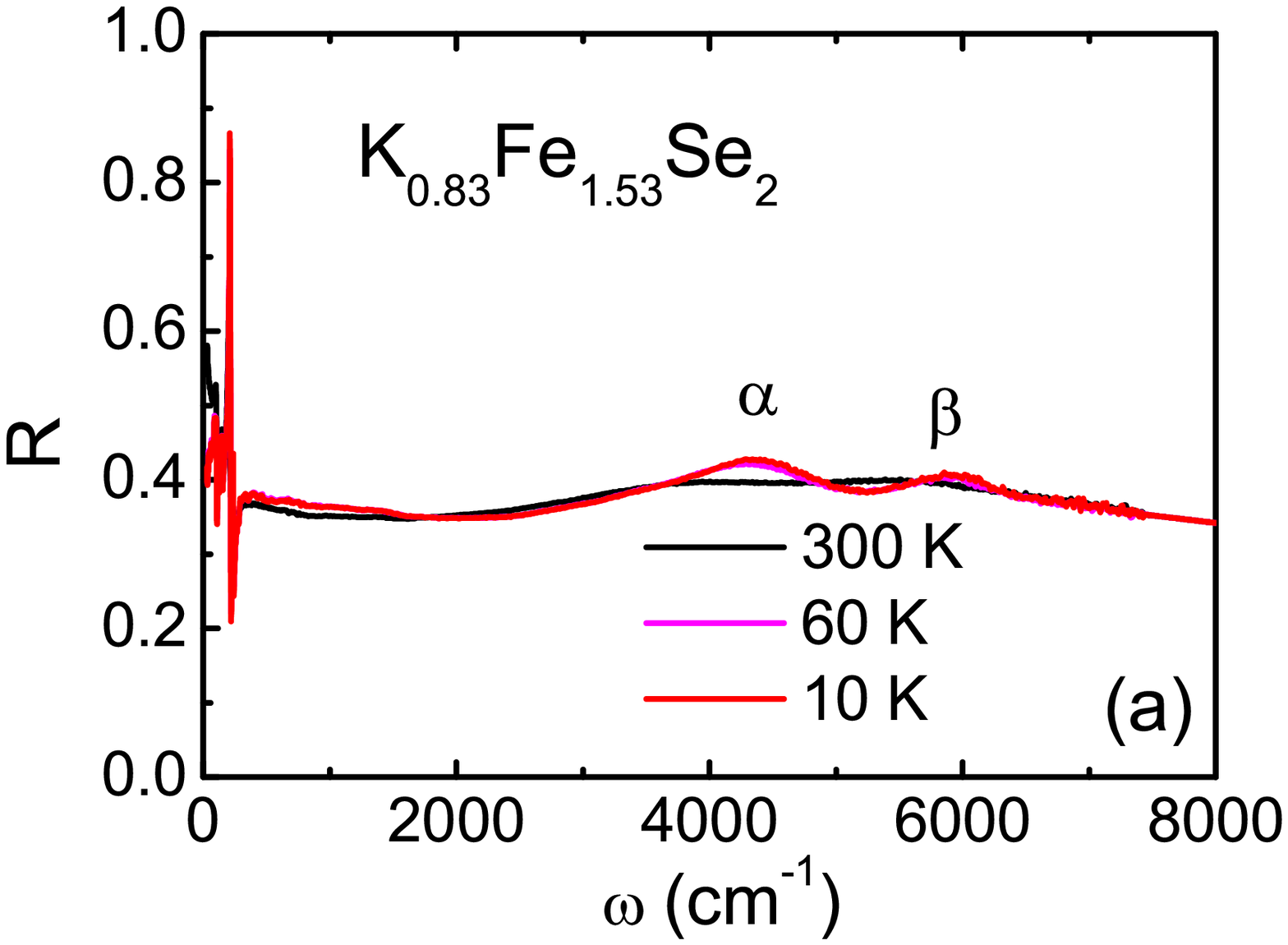}
\includegraphics[clip,width=1.65in]{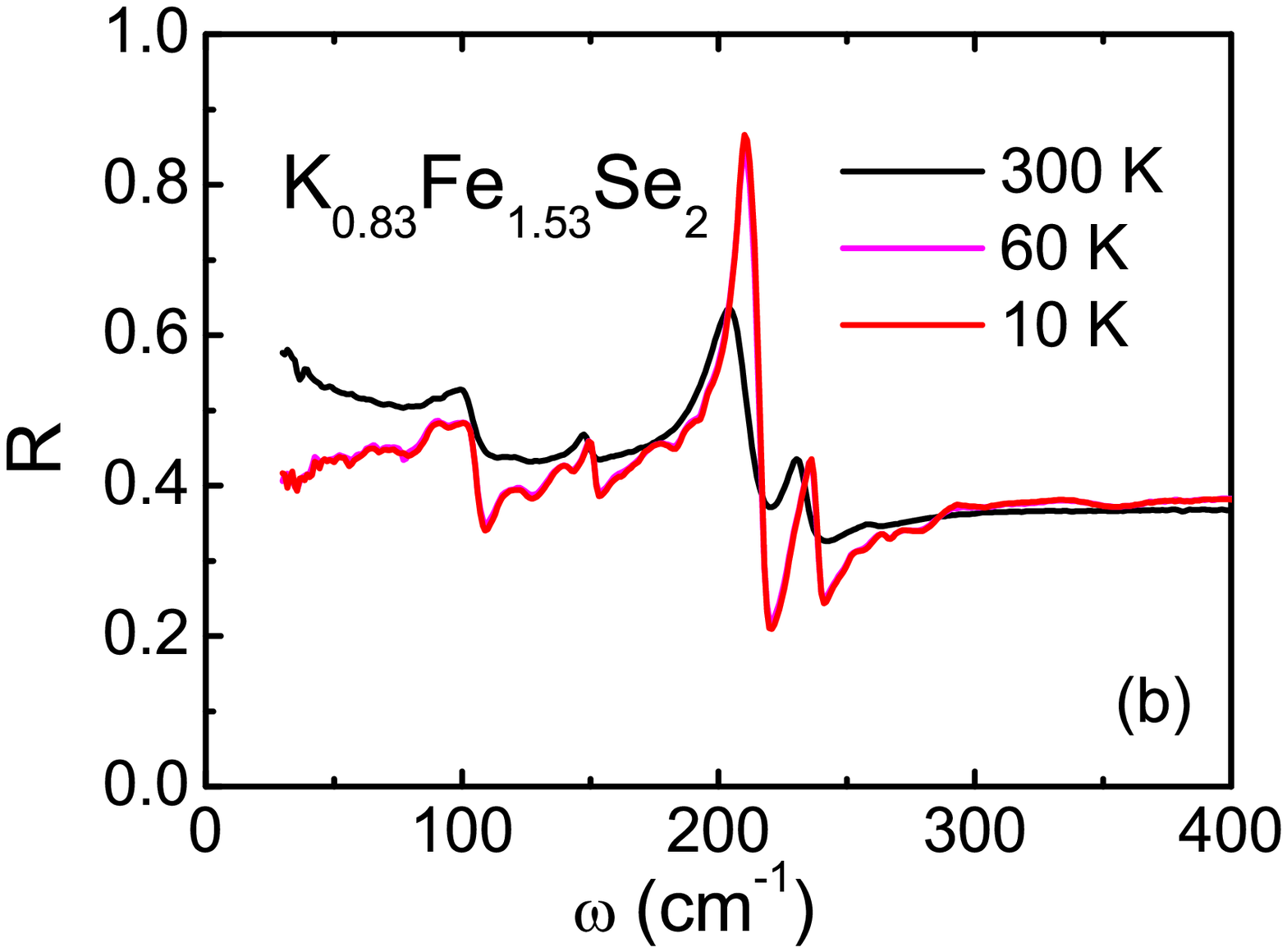}
\includegraphics[clip,width=1.65in]{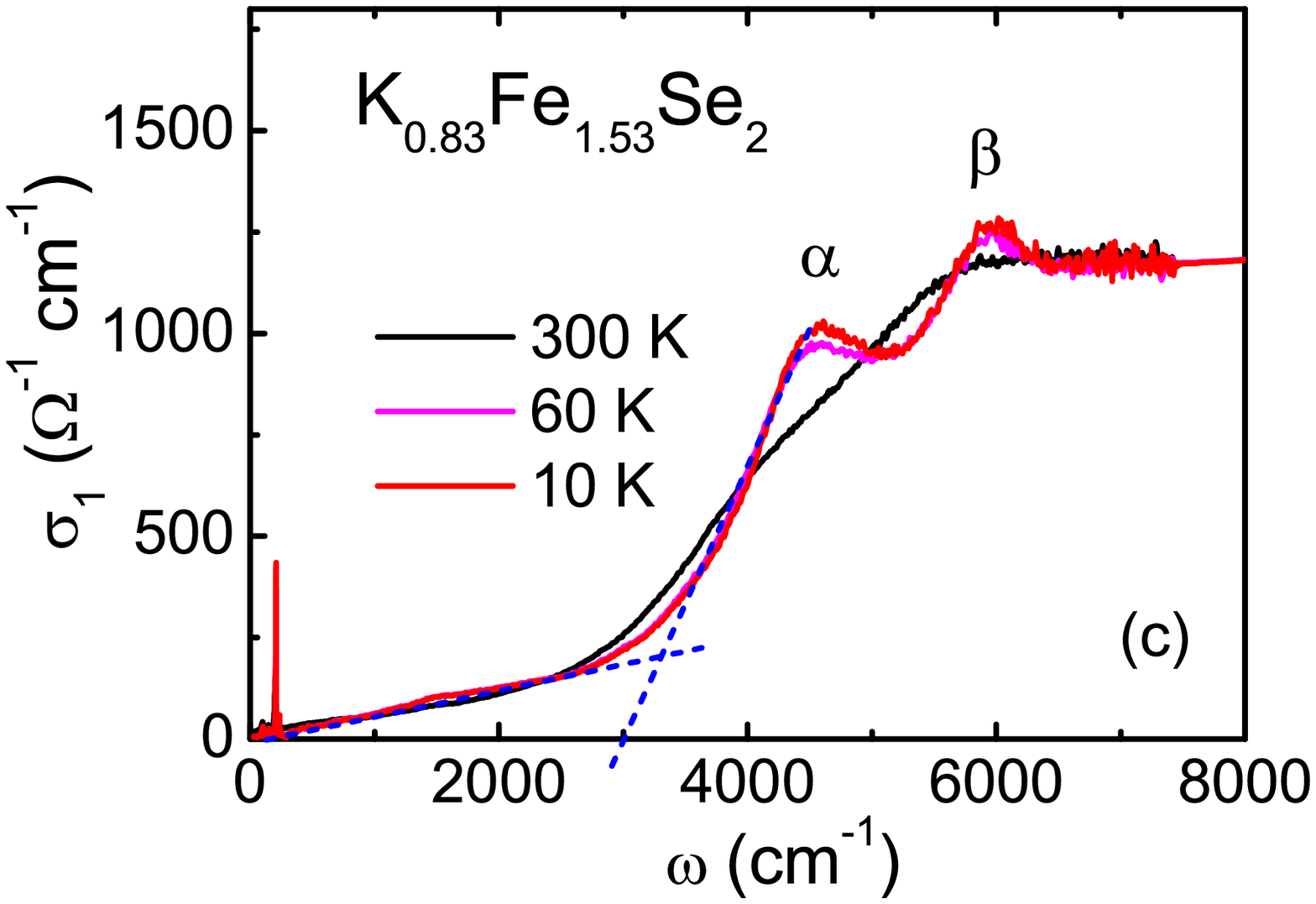}
\includegraphics[clip,width=1.65in]{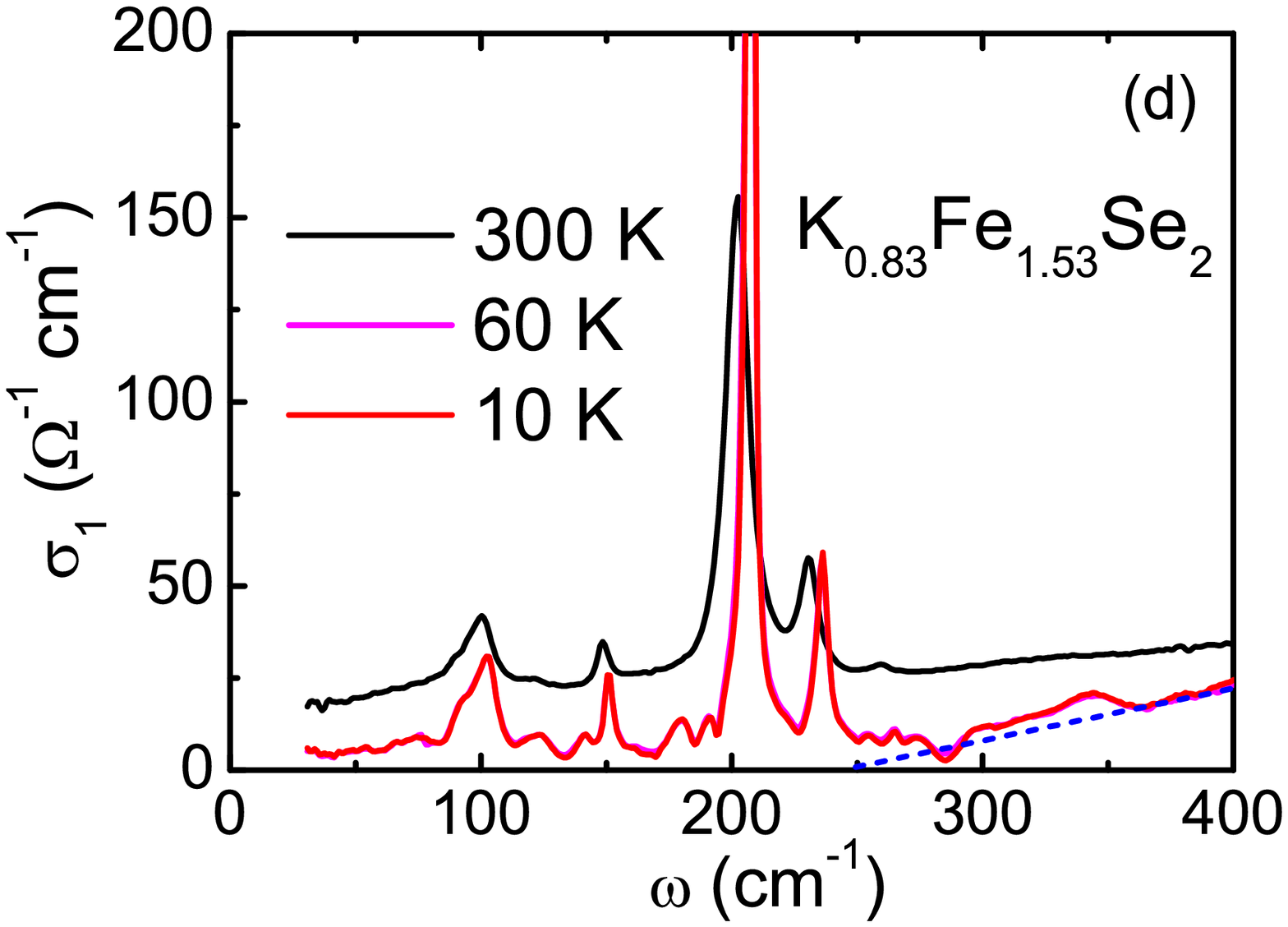}
\caption{(Color online) (a) optical reflectance R($\omega$)
spectra at different temperatures up to 8000 \cm. (b) An expanded
plot of R($\omega$) spectra below 400 \cm. (c) the conductivity
$\sigma_1(\omega)$ spectra of the sample at different temperatures
up to 8000 \cm. (d) An expanded plot of $\sigma_1$($\omega$)
spectra below 400 \cm. The dashed straight lines are used for
extrapolation of energy gaps.}
\end{figure}

Besides the semiconducting/insulating response, there exist two
prominent features in the spectra. First, the two-peaks interband
transition structure in the mid- to near infrared region between
$\sim$ 4500 \cm and 6000 \cm seen at room temperature become very
prominent at low temperature, as shown clearly in Fig. 3 (a) and
(c). Second, as evidenced clearly in Fig. 2 (b) and (d), there
exist over 10 phonon peaks at the low frequencies. It should be
noted, for the standard 122 structure, only two infrared active
phone modes should be present in the ab-plane infrared
spectra\cite{Akrap}.

As the superconductivity is in close proximity to the insulating
phase, it is extremely important to identify the properties and
nature of this insulating compound. A crucial question is whether
the compound is a Mott insulator or a band insulator? This would
be a starting point for understanding the material. Actually,
there have been controversial points of views on the basis of
weak-coupling or strong coupling approaches in understanding the
magnetism and superconductivity in Fe-pnictides/chalcogenides ever
since the discovery of Fe-based superconductors.

From Fig. 3 (c) and (d), we could see that the low-$\omega$
$\sigma_1(\omega)$ at 10 K is dominated by the phonon peaks with a
vanishing electronic contribution roughly below 250 \cm. Above
this energy the conductivity increases almost linearly up to 3000
\cm at which much stronger interband transitions appear. Based on
this direct observation, we can assign the sharp increase near
3000 \cm ($\sim$0.37 eV) in $\sigma_1(\omega)$ as a gap arising
from a direct interband transition. However, below 3000 \cm, there
still exist sizeable absorptions, most likely from an indirect
interband transitions assisted by the impurities and collective
boson excitations. But this absorption energy scale is much
smaller, $\sim$ 250 \cm (30 meV) as obtained from a linear
extrapolation of the $\sigma_1(\omega)$ below 3000 \cm. Since the
gap (even for the direct energy gap) is small, the compound is not
likely to be a Mott insulator, given the fact that the Mott gap
should have an energy scale comparable to the on-site Coulomb
repulsion energy U. The indirect nature of the interband
transition for the lowest gap in the electronic excitations
appears also to rule out the possibility of the presence of a
charge transfer energy gap as a case established for the case of
high-T$_c$ cuprates. We suggest that the compound should be
considered as a small band gap semiconductor. Recent band
structural calculations \cite{CaoDai,Yan} could indeed reproduce
the semiconducting band structures by taking account of the
ordering of Fe vacancies with the experimentally determined
blocked checkerboard AFM order although the calculated band gaps
are larger, $\sim$0.4-0.6 eV. It is also in agreement with recent
ARPES measurement\cite{Qian} on a superconducting sample where the
band renormalization factor is found to be only 2.5.

\begin{figure}
\includegraphics[width=7cm]{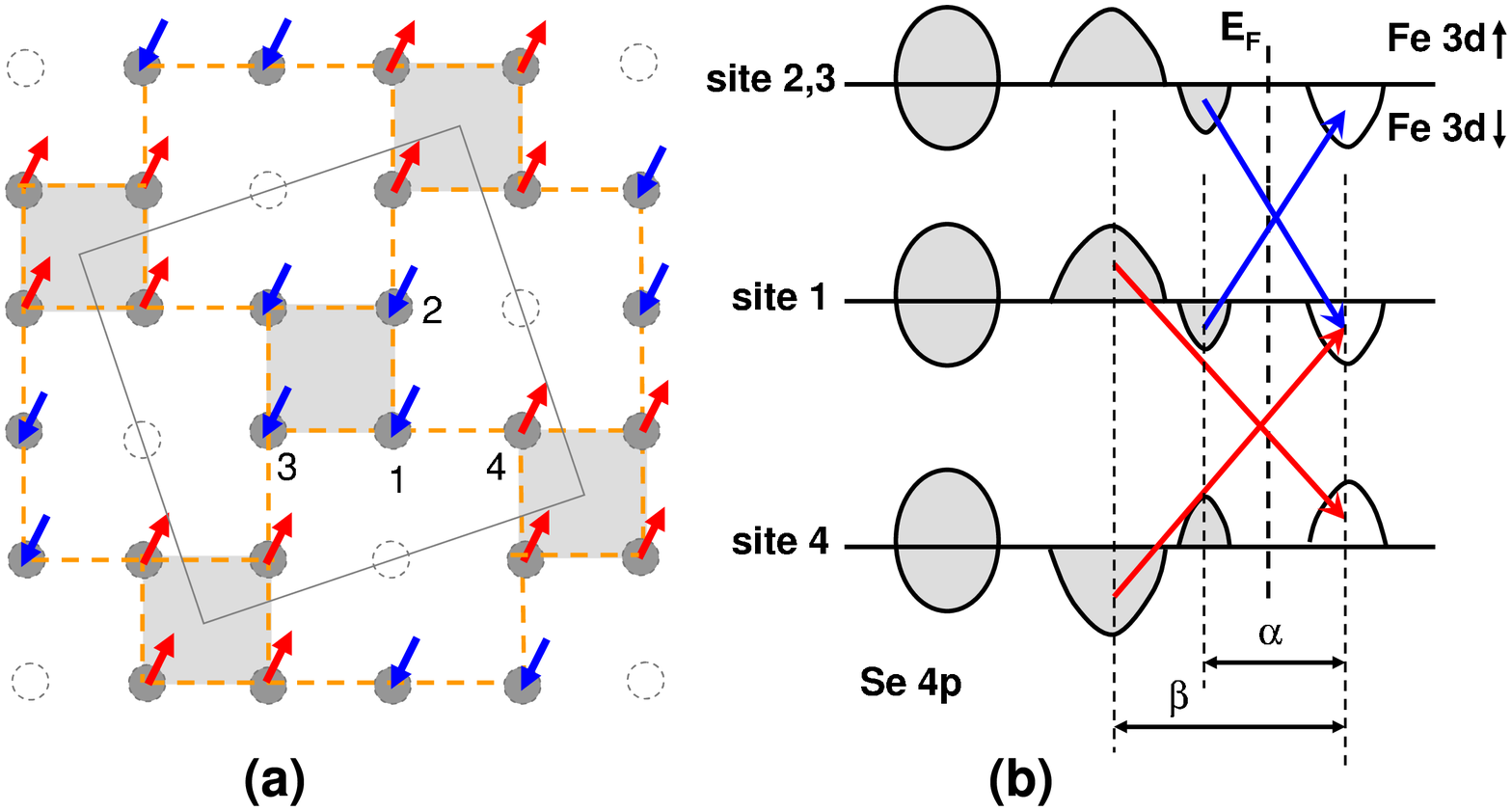}
\caption{(Color online) (a) The magnetic structure determined by
the neutron diffraction experiment \cite{Bao1}. The grey solid
lines represent a $\sqrt{5}\times\sqrt{5}\times1$ unit cell. (b) A
schematic picture for the band occupation for each Fe site. Site 4
has opposite spin-polarized occupations with site 1, 2, and 3.
Possible d-d transitions are indicated by the arrows. The double
peaks represents the coexistence of the ferro- and
antiferromagnetic correlations between the neighboring sites.}
\end{figure}

As mentioned above, the double interband transition peaks in the
mid- to near-infrared region and abundant phonon modes below 400
\cm are two other interesting observations. They were not seen in
any other Fe-pnictides/chalcogenides, and therefore, could be
taken as the characteristic optical spectral features for this
specific system. We believe that both should be strongly
associated with the Fe vacancy ordering in the compound. We shall
first address the origin of the double peak structure. For the
investigated system A$_x$Fe$_{2-y}$Se$_2$, Fe vacancies obviously
exist. Those Fe vacancies should form ordered pattern otherwise it
is difficult to imagine that superconductivity could emerge in
this system. For the composition of Fe$\sim$1.5, one expects to
see one vacancy per four Fe sites. Several different Fe vacancy
ordering patterns matching with one vacancy per four sites have
been proposed \cite{Fang,Yan2,YuSi}, they all lead to inequivalent
Fe sites in the structure with different coordinated Fe neighbors.
This may cause energy difference between bonding and antibonding
states of Fe 3d orbitals, which may explain the observed double
peaks structure. However, this possibility is essentially ruled
out by the neutron diffraction experiments \cite{Bao1,FYe,Bao2},
which indicated a complete absence of an ordering pattern with one
vacancy per four Fe sites. Instead, for all available
A$_x$Fe$_{2-y}$Se$_2$ (A=K, Rb, Cs, Tl) samples with different Fe
compositions, neutron measurements \cite{Bao2} revealed a Fe
vacancy ordering with a $\sqrt{5}\times\sqrt{5}\times1$
superlattice pattern below room temperature. In this ordering
pattern, all occupied Fe sites are equivalent. The major
difference for different Fe compositions would be a change of
occupation rates for both occupied and vacant Fe sites.

We find that the $\alpha$ and $\beta$ double peaks could be
naturally explained by the peculiar magnetic structure (shown in
Fig. 4 (a)) determined by the neutron diffraction experiment
\cite{Bao1,FYe,Bao2}. In this ordering pattern, each Fe site is
ferromagnetic coupled with two neighboring Fe sites and
antiferromagnetic coupled with the third neighboring Fe site. To
simplify our discussions, we consider that the Fe ion has a 2+
valance state with 3d$^6$ configuration. We ignore possible
itinerant d electrons from certain orbitals, and assume that the
Hund's rule coupling energy is larger than the crystal field
splitting energy, then the 5 electrons would fully occupy 5
orbitals in a spin polarized direction, the other one electron
would occupy one of the orbital in the opposite spin direction,
leaving the other four orbitals completely unoccupied. The band
occupation for each Fe site is schematically shown in Fig. 4 (b).
In Fig. 4 (a), the Fe 1, Fe2, and Fe3 have the same spin-polarized
occupations, the Fe 4 has the opposite spin-polarized occupations.
Because the optical transitions do not involve spin-flip process,
the possible d-d transitions (via hybridization with Se 4p
orbitals) are indicated by the arrows, representing the
transitions from the occupied states at one Fe site to the
unoccupied states at nearest-neighboring Fe sites. From this
schematic diagram, we can see that the $\alpha$ peak comes from
the transitions between the neighboring ferromagnetic ordered
sites and the $\beta$ transition from the neighboring
antiferromagnetic ordered sites. The double peaks represents the
coexistence of the ferro- and antiferromagnetic correlations
between the neighboring sites.

Another striking feature is the presence of more than 10 phonon
peaks at the low frequencies. As we also mentioned, for the
standard 122 structure, only two infrared active phone modes
should be present in the ab-plane infrared spectra\cite{Akrap}. As
the neutron measurements indicate the formation of
$\sqrt{5}\times\sqrt{5}\times1$ superlattice pattern due to the Fe
vacancy ordering, the crystallographic unit cell is actually
enlarged as K$_2$Fe$_4$Se$_5$.\cite{Bao1} The composition
K$_{0.83}$Fe$_{1.53}$Se$_2$ deviates from this perfect 245
compound, leading to a modified occupation rates in the occupied
and vacant Fe sites (or disorders) based on the neutron
diffraction measurements. Then more phonon modes than the
tetragonal ThCr$_2$Si$_2$ structure would be allowed in optical
measurement.

To conclude, we performed infrared spectroscopy study on
semiconducting of K$_{0.83}$Fe$_{1.53}$Se$_2$. We identify that
the compound is a small energy gap semiconductor. Our infrared
measurement also revealed two other characteristic spectral
features which are specific to the materials but absent in other
Fe-pnictides/chalcogenides: a double peak structure near 4000-6000
\cm and abundant phonon modes much more than those expected for a
standard 122 structure. We elaborated that both could be naturally
explained from the peculiar blocked AFM structure due to the
presence of Fe vacancy ordering.

We acknowledge helpful discussions with Z. Fang, D. Xi, G. M.
Zhang, H. Ding, Z. Y. Lu, T. Xiang, M. F. Fang, G. F. Chen, Q. Si
and Y. P. Wang. This work was supported by the NSFC, CAS and the
973 project of the MOST.

%\bibliography{MgIrB}% Produces the bibliography vi BibTeX.

\end{document}